\documentclass[12pt]{article}
\usepackage[pctex32]{epsfig}
\newcommand{\nc}{\newcommand}
\nc{\renc}{\renewcommand}

\nc{\half}{{\textstyle{1\over2}}}

\nc{\etal}{\mbox{\it et al. }}
\nc{\ie}{{\it i.e.}}
\nc{\eg}{{\it e.g.}}

\renc{\thefootnote}{\arabic{footnote}}
\nc{\capt}[1]{{\bf Figure.} {\small\sl #1}}

\nc{\eqs}[2]{\mbox{Eqs.~(\ref{#1},\,\ref{#2})}}
\nc{\eq}[1]{\mbox{Eq.~(\ref{#1})}}

\nc{\figs}[2]{\mbox{Figs.~(\ref{#1},\,\ref{#2})}}
\nc{\fig}[1]{\mbox{Fig~.(\ref{#1})}}

\nc{\tag}[1]{\label{#1} \marginpar{{\footnotesize #1}}}
\nc{\mtag}[1]{\label{#1} \mbox{\marginpar{{\footnotesize #1}}}}
\renc{\baselinestretch}{1.2}
\newlength{\overeqskip}
\newlength{\undereqskip}
\nc{\be}[1]{\begin{equation} \mbox{$\label{#1}$}}
\nc{\bea}[1]{\begin{eqnarray} \mbox{$\label{#1}$}}
\nc{\Section}[2]{\section{#2}\label{#1}}
\nc{\Bibitem}[1]{\bibitem{#1}}
\nc{\Label}[1]{\label{#1}}

\nc{\eea}{\vspace{\undereqskip}\end{eqnarray}}
\nc{\ee}{\vspace{\undereqskip}\end{equation}}
\nc{\bdm}{\begin{displaymath}}
\nc{\edm}{\end{displaymath}}
\nc{\dpsty}{\displaystyle}
\nc{\bc}{\begin{center}}
\nc{\ec}{\end{center}}
\nc{\ba}{\begin{array}}
\nc{\ea}{\end{array}}
\nc{\bab}{\begin{abstract}}
\nc{\eab}{\end{abstract}}
\nc{\btab}{\begin{tabular}}
\nc{\etab}{\end{tabular}}
\nc{\bit}{\begin{itemize}}
\nc{\eit}{\end{itemize}}
\nc{\ben}{\begin{enumerate}}
\nc{\een}{\end{enumerate}}
\nc{\bfig}{\begin{figure}}
\nc{\efig}{\end{figure}}
\nc{\arreq}{&\!=\!&}
\nc{\arrmi}{&\!-\!&}
\nc{\arrpl}{&\!+\!&}
\nc{\arrap}{&\!\!\!\approx\!\!\!&}
\nc{\non}{\nonumber\\*}
\nc{\align}{\!\!\!\!\!\!\!\!&&}

\def\lsim{\; \raise0.3ex\hbox{$<$\kern-0.75em
      \raise-1.1ex\hbox{$\sim$}}\; }
\def\gsim{\; \raise0.3ex\hbox{$>$\kern-0.75em
      \raise-1.1ex\hbox{$\sim$}}\; }
\nc{\DOT}{\hspace{-0.08in}{\bf .}\hspace{0.1in}}
\nc{\Laada}{\hbox {$\sqcap$ \kern -1em $\sqcup$}}
\nc\loota{{\scriptstyle\sqcap\kern-0.55em\hbox{$\scriptstyle\sqcup$}}}
\nc\Loota{{\sqcap\kern-0.65em\hbox{$\sqcup$}}}
\nc\laada{\Loota}
\nc{\qed}{\hskip 3em \hbox{\BOX} \vskip 2ex}

\nc{\real}{{\rm I \! R}}
\nc{\Z}{{\sf Z \!\!\! Z}}
\nc{\complex}{{\rm C\!\!\! {\sf I}\,\,}}
\def\bigid{\leavevmode\hbox{\small1\kern-3.8pt\normalsize1}}
\def\id{\leavevmode\hbox{\small1\kern-3.3pt\normalsize1}}
\nc{\slask}{\!\!\!/}
\nc{\bis}{{\prime\prime}}
\nc{\pa}{\partial}
\nc{\na}{\nabla}
\nc{\ra}{\rangle}
\nc{\la}{\langle}
\nc{\goto}{\rightarrow}
\nc{\swap}{\leftrightarrow}

\nc{\EE}[1]{ \mbox{$\cdot10^{#1}$} }
\nc{\abs}[1]{\left|#1\right|}
\nc{\at}[2]{\left.#1\right|_{#2}}
\nc{\norm}[1]{\|#1\|}
\nc{\abscut}[2]{\Abs{#1}_{\scriptscriptstyle#2}}
\nc{\vek}[1]{{\rm\bf #1}}
\nc{\integral}[2]{\int\limits_{#1}^{#2}}
\nc{\inv}[1]{\frac{1}{#1}}
\nc{\dd}[2]{{{\partial #1}\over{\partial #2}}}
\nc{\ddd}[2]{{{{\partial}^2 #1}\over{\partial {#2}^2}}}
\nc{\dddd}[3]{{{{\partial}^2 #1}\over
        {\partial #2 \partial #3}}}
\nc{\dder}[2]{{{d #1}\over{d #2}}}
\nc{\ddder}[2]{{{d^2 #1}\over{d {#2}^2}}}
\nc{\dddder}[3]{{d^2 #1}\over
        {d #2 d #3}}
\nc{\dx}[1]{d\,^{#1}x}
\nc{\dy}[1]{d\,^{#1}y}
\nc{\dz}[1]{d\,^{#1}z}
\nc{\dl}[1]{\frac{d\,^{#1}l}{(2\pi)^{#1}}}
\nc{\dk}[1]{\frac{d\,^{#1}k}{(2\pi)^{#1}}}
\nc{\dq}[1]{\frac{d\,^{#1}q}{(2\pi)^{#1}}}

\nc{\cc}{\mbox{$c.c.$ }}
\nc{\hc}{\mbox{$h.c.$ }}
\nc{\cf}{cf.\ }
\nc{\erfc}{{\rm erfc}}
\nc{\Tr}{{\rm Tr\,}}
\nc{\tr}{{\rm tr\,}}
\nc{\pol}{{\rm pol}}
\nc{\sign}{{\rm sign}}
\nc{\bfT}{{\bf T }}

\nc{\cA}{{\cal A}}
\nc{\cB}{{\cal B}}
\nc{\cD}{{\cal D}}
\nc{\cE}{{\cal E}}
\nc{\cG}{{\cal G}}
\nc{\cH}{{\cal H}}
\nc{\cL}{{\cal L}}
\nc{\cO}{{\cal O}}
\nc{\cT}{{\cal T}}
\nc{\cN}{{\cal N}}
\nc{\rvac}[1]{|{\cal O}#1\rangle}
\nc{\lvac}[1]{\langle{\cal O}#1|}
\nc{\rvacb}[1]{|{\cal O}_\beta #1\rangle}
\nc{\lvacb}[1]{\langle{\cal O}_\beta #1 |}
\nc{\bb}{\bar{\beta}}
\nc{\bt}{\tilde{\beta}}
\nc{\ctH}{\tilde{\cal H}}
\nc{\chH}{\hat{\cal H}}
\nc{\1}{\aa}
\nc{\2}{\"{a}}
\nc{\3}{\"{o}}
\nc{\4}{\AA}
\nc{\5}{\"{A}}
\nc{\6}{\"{O}}
\nc{\al}{\alpha}
\nc{\g}{\gamma}
\nc{\Del}{\Delta}
\nc{\e}{\epsilon}
\nc{\eps}{\epsilon}
\nc{\lam}{\lambda}
\nc{\om}{\omega}
\nc{\Om}{\Omega}
\nc{\ve}{\varepsilon}
\nc{\mn}{{\mu\nu}}
\nc{\vp}{\varphi}

\nc{\advp}[3]{{\it  Adv.\ in\ Phys.\ }{{\bf #1} {(#2)} {#3}}}
\nc{\annp}[3]{{\it  Ann.\ Phys.\ (N.Y.)\ }{{\bf #1} {(#2)} {#3}}}
\nc{\apl}[3]{{\it  Appl. Phys. Lett. }{{\bf #1} {(#2)} {#3}}}
\nc{\apj}[3]{{\it  Ap.\ J.\ }{{\bf #1} {(#2)} {#3}}}
\nc{\apjl}[3]{{\it  Ap.\ J.\ Lett.\ }{{\bf #1} {(#2)} {#3}}}
\nc{\aap}[3]{{\it Astron.\ Astrophys.\ }{{\bf #1} {(#2)} {#3}}}
\nc{\app}[3]{{\it Astropart.\ Phys.\ }{{\bf #1} {(#2)} {#3}}}
\nc{\cmp}[3]{{\it  Comm.\ Math.\ Phys.\ }{{ \bf #1} {(#2)} {#3}}}
\nc{\cqg}[3]{{\it  Class.\ Quant.\ Grav.\ }{{\bf #1} {(#2)} {#3}}}
\nc{\epl}[3]{{\it  Europhys.\ Lett.\ }{{\bf #1} {(#2)} {#3}}}
\nc{\ijmp}[3]{{\it Int.\ J.\ Mod.\ Phys.\ }{{\bf #1} {(#2)} {#3}}}
\nc{\ijtp}[3]{{\it Int.\ J.\ Theor.\ Phys.\ }{{\bf #1} {(#2)} {#3}}}
\nc{\jmp}[3]{{\it  J.\ Math.\ Phys.\ }{{ \bf #1} {(#2)} {#3}}}
\nc{\jpa}[3]{{\it  J.\ Phys.\ A\ }{{\bf #1} {(#2)} {#3}}}
\nc{\jpc}[3]{{\it  J.\ Phys.\ C\ }{{\bf #1} {(#2)} {#3}}}
\nc{\jap}[3]{{\it J.\ Appl.\ Phys.\ }{{\bf #1} {(#2)} {#3}}}
\nc{\jpsj}[3]{{\it J.\ Phys.\ Soc.\ Japan\ }{{\bf #1} {(#2)} {#3}}}
\nc{\lmp}[3]{{\it Lett.\ Math.\ Phys.\ }{{\bf #1} {(#2)} {#3}}}
\nc{\mpl}[3]{{\it  Mod.\ Phys.\ Lett.\ }{{\bf #1} {(#2)} {#3}}}
\nc{\ncim}[3]{{\it  Nuov.\ Cim.\ }{{\bf #1} {(#2)} {#3}}}
\nc{\np}[3]{{\it  Nucl.\ Phys.\ }{{\bf #1} {(#2)} {#3}}}
\nc{\pan}[3]{{\it Phys.\ Atom.\ Nucl. }{{\bf #1} {(#2)} {#3}}}
\nc{\pr}[3]{{\it Phys.\ Rev.\ }{{\bf #1} {(#2)} {#3}}}
\nc{\pra}[3]{{\it  Phys.\ Rev.\ A\ }{{\bf #1} {(#2)} {#3}}}
\nc{\prb}[3]{{\it  Phys.\ Rev.\ B\ }{{{\bf #1} {(#2)} {#3}}}}
\nc{\prc}[3]{{\it  Phys.\ Rev.\ C\ }{{\bf #1} {(#2)} {#3}}}
\nc{\prd}[3]{{\it  Phys.\ Rev.\ D\ }{{\bf #1} {(#2)} {#3}}}
\nc{\prl}[3]{{\it Phys.\ Rev.\ Lett.\ }{{\bf #1} {(#2)} {#3}}}
\nc{\pl}[3]{{\it  Phys.\ Lett.\ }{{\bf #1} {(#2)} {#3}}}
\nc{\prep}[3]{{\it Phys.\ Rep.\ }{{\bf #1} {(#2)} {#3}}}
\nc{\prsl}[3]{{\it Proc.\ R.\ Soc.\ London\ }{{\bf #1} {(#2)} {#3}}}
\nc{\ptp}[3]{{\it  Prog.\ Theor.\ Phys.\ }{{\bf #1} {(#2)} {#3}}}
\nc{\ptps}[3]{{\it  Prog\ Theor.\ Phys.\ suppl.\ }{{\bf #1} {(#2)} {#3}}}
\nc{\physa}[3]{{\it  Physica\ A\ }{{\bf #1} {(#2)} {#3}}}
\nc{\physb}[3]{{\it  Physica\ B\ }{{\bf #1} {(#2)} {#3}}}
\nc{\phys}[3]{{\it Physica\ }{{\bf #1} {(#2)} {#3}}}
\nc{\rmp}[3]{{\it  Rev.\ Mod.\ Phys.\ }{{\bf #1} {(#2)} {#3}}}
\nc{\rpp}[3]{{\it Rep.\ Prog.\ Phys.\ }{{\bf #1} {(#2)} {#3}}}
\nc{\sjnp}[3]{{\it Sov.\ J.\ Nucl.\ Phys.\ }{{\bf #1} {(#2)} {#3}}}
\nc{\spjetp}[3]{{\it Sov.\ Phys.\ JETP\ }{{\bf #1} {(#2)} {#3}}}
\nc{\yf}[3]{{\it Yad.\ Fiz.\ }{{\bf #1} {(#2)} {#3}}}
\nc{\zetp}[3]{{\it Zh.\ Eksp.\ Teor.\ Fiz.\  }{{\bf #1}  {(#2)} {#3}}}
\nc{\zp}[3]{{\it Z.\ Phys.\ }{{\bf #1} {(#2)} {#3}}}
\nc{\ibid}[3]{{\sl ibid.\ }{{\bf #1} {#2} {#3}}}
\nc{\rf}[1]{(\ref{#1})}
\nc{\nn}{\nonumber \\*}
\nc{\bfB}{\bf{B}}
\nc{\bfv}{\bf{v}}
\nc{\bfx}{\bf{x}}
\nc{\bfy}{\bf{y}}
\nc{\vx}{\vec{x}}
\nc{\vy}{\vec{y}}
\nc{\oB}{\overline{B}}
\nc{\oI}{\overline{I}}
\nc{\oR}{\overline{R}}
\nc{\rar}{\rightarrow}
\nc{\ti}{\times}
\nc{\slsh}{\hskip-5pt/}
\nc{\sm}{Standard~Model~}
\nc{\MP}{M_{\rm Pl}}
\nc{\tp}{t_{\rm Pl}}
\nc{\ave}{\bar{E}}

\nc{\eff}{{\rm eff}}
\nc{\kk}{\vek{k}}
\nc{\pp}{{\rm p}}
\nc{\ga}{g_{a\gamma}}
\nc{\vv}{\\}
\nc{\eee}{{\bf E}}
\nc{\bbb}{{\bf B}}
\nc{\qcd}{T_{\rm QCD}}
\nc{\G}{\rm \ G}
\def\vec#1{{\bf #1}}
\begin{document}
{\title{\vskip-2truecm{\hfill {{\small HIP-1998-35/TH\\
        }}\vskip 1truecm}
{\bf Testing neutrino magnetic moments with AGNs}}


{\author{
{\sc Kari Enqvist$^{1}$}\\ 
{\sl\small Department of Physics and Helsinki Institute of Physics}
\\ {\sl\small P.O. Box 9, FIN-00014 University of Helsinki,
Finland}\\
{\sc Petteri Ker\"anen$^{2}$ and Jukka
Maalampi$^{3}$  }\\
{\sl\small Department of Physics, P.O. Box 9,
FIN-00014 University of Helsinki,
Finland}
}
\maketitle
\vspace{2cm}
\begin{abstract}
\noindent
We propose to test the magnetic transition moments of Majorana
neutrinos by comparing the fluxes of different flavours of neutrinos
coming from active galactic nuclei (AGN). We show that, with reasonable assumptions about
the magnetic field of the AGN, it is possible to obtain limits on
$\nu_{\tau}\nu_{e}$ and $\nu_{\tau}\nu_{\mu}$ transition moments
which are three to five orders of magnitude better than the
laboratory limits. We also point out that with certain parameter values
the ratio $\nu_{\tau}/\nu_{e,\mu}$, when measured from different sources, 
is expected to vary from zero to values somewhat higher than one, providing
an unambigious signal of a magnetic transition within the AGN which cannot be explained by 
neutrino oscillations.
\\
\\
\noindent
{\it PACS:} 13.15.+g; 14.60.St; 98.54.Cm; 98.70.Sa

\noindent
{\it Keywords:} active galactic nuclei; neutrino magnetic moments
\end{abstract}
\vfil
\footnoterule
{\small 
$^{1}$kari.enqvist@helsinki.fi;
$^{2}$petteri.keranen@helsinki.fi;
$^{3}$jukka.maalampi@helsinki.fi
\thispagestyle{empty}
\newpage
\setcounter{page}{1}


\noindent {\it 1. Introduction.} 
Recent Super-Kamiokande~\cite{kamioka} results imply that neutrinos have a non-vanishing
mass  and therefore also a non-vanishing magnetic moment.
As a consequence, the wave function of a neutrino traversing in a
magnetic field will be subject to spin rotation. For a large enough
magnetic moment this effectively results in a helicity flip. If
neutrinos are Dirac particles, helicity flip induces a conversion of an
active left-handed state
$\nu_{iL}$ into a sterile right-handed state $\nu_{jR}$, thereby
depleting the measurable neutrino flux. Majorana neutrinos do not have
diagonal magnetic moments but  may have non-diagonal transition
magnetic moments which give rise to
$N_{iL}\rightarrow N_{jR} \ (i \neq j)$ conversions. In contrast with the
Dirac case, the right-handed state $N_{jR}$ is here active and detectable
$(N_{jR}\sim \nu^c_{jR})$, and hence the measurable neutrino flux is not
depleted in the Majorana case, only its composition is changed.
This is of particular interest for the ultrahigh energy neutrino flux
emitted by active
galactic nuclei (AGN), which are believed to be powered
by black holes and sustain large magnetic fields to accelerate protons,
which through collisions produce neutrinos 
\cite{Gaisser}. It is therefore possible that magnetic
moment induced helicity flips could affect the intensity and/or
composition of the AGN neutrino flux measured in detectors on Earth,
such as AMANDA, Nestor, Baikal and ANTARES. 

For the purposes of the present study, there are two types of sources of
interest, blazars  and hot spots.
Magnetic fields in blazar jets are assumed be of the order of 1 G and in
hot spots some fraction of mG~\cite{Gaisser,RachenBiermann}. The
characteristic size of these objects is of the order of $10^{-2}\ \rm pc$
and 1 kpc, respectively, and as we shall show, then magnetic
moments of the order of $10^{-15}-10^{-14} \mu_B$ would be large enough
to cause detectable effects. Intergalactic magnetic fields are too
weak to induce any additional transitions, the estimated upper limit
for their field strength being about \cite{Mag} $10^{-10}\ \rm G$.
Galactic magnetic fields have a strength of the order of
$10^{-6}\ \rm G$ \cite{galaxy} but have much smaller spatial dimensions
and can be neglected as well in first approximation.

The most unambigious signal of a magnetic transition would be the
appearance of a tau neutrino component in the neutrino flux. The tau
neutrinos
$\nu_\tau$, in contrast  with $\nu_e$ and $\nu_\mu$, are not produced in
any substantial amounts in the processes inside the source, but they may
be generated by magnetic interactions. Therefore
in the following we will focus on the appearance
of tau neutrinos. We assume that neutrinos are Majorana particles
with non-vanishing transition magnetic moments that allow
$\nu_e\rightarrow\nu_\tau$ and $\nu_\mu\rightarrow\nu_\tau$ transitions.
The magnetic flavor transitions would result in a neutrino spectrum
different from the expected one.

There are two suggested signatures for tau neutrino detection in neutrino
telescopes: the so-called
double bang event, and the absence of absorption 
by the Earth. In a double bang event~\cite{LearnedPakvasa} the tau
neutrino, interacting 
with a nucleus via $W$-boson exchange in the detector, will produce a tau
lepton and a 
hadronic jet and the decaying tau lepton will, if its energy is
around 1 PeV, 
produce another jet of particles inside the detector. The energy and
direction of the tau neutrino can be
quite well defined if two practically same timed jets are tagged.
The absence of absorption, a more recent idea to detect tau
neutrinos~\cite{HalzenSaltz},
is based on the opacity of the Earth at high neutrino energies~\cite{Gandhi}.
All neutrino species
scatter from matter, but in the case of the tau neutrino there is always a less
energetic $\nu_\tau$ from the decaying tau lepton in 
the final state. When propagating through the Earth the sustained tau neutrino component will
lose its energy in sequential scattering processes. Finally 
the energy will be low enough and the neutrinos will not interact any longer, resulting in an
excess of upgoing events at energies around 10 - 100 TeV. 

As mentioned above, the AGN neutrinos are sensitive to magnetic moment
values of the order of $10^{-15}-10^{-14} \mu_B$. This is an interesting
region since such small values cannot be tested in laboratory
experiments, nor completely by traditional astrophysics or
cosmology.
Laboratory limits for the magnetic moments of the different neutrino
species are~\cite{Derbin,Krakauer}
\bea{limits}
\mu_{\nu_e} < 1.8\times 10^{-10} \mu_B, \\
\mu_{\nu_\mu} < 7.4\times 10^{-10}\mu_B.
\eea
These limits are based on elastic scattering cross sections and
are therefore valid for both Dirac neutrino helicity flips and Majorana
neutrino magnetic transition moments.   The most stringent
astrophysical limit,
$\mu \leq 1.4\times 10^{-12}
\mu_B$, is based on the cooling rates of red giants~\cite{Raffelt} and
is valid for all Majorana neutrino flavors. Limits from the energy loss
of SN1987A are valid for Dirac neutrinos only. Nucleosynthesis
constraints~\cite{Morgan} are of the order of $10^{-11} \mu_B$ and
apply also to Majorana neutrinos. 
\\

\noindent{\it 2. Neutrino production and magnetic conversion in the sources.} 
In blazars protons are accelerated in jets of plasma~\cite{MaBie}
bursting out along the rotation axis of the central engine, a supermassive
black hole. Hot spots are created when a jet collides with plasma and gas
of a lobe. They may be even around 1 Mpc away from the center of the host
galaxy. 
In the sources high energy protons are believed to be accelerated by the
first order Fermi acceleration mechanism by repeated scatterings back and
forth accross a shock front in a partially turbulent magnetic field. The
gyroradius of the proton will grow as the kinetic energy
increases~\cite{accel}. This process can go on untill the particle either
escapes the area of acceleration or scatters from another particle.
Electron and muon neutrinos are produced in pion photoproduction process,
the most important chain being $p\gamma \rightarrow
n\pi^+\rightarrow\mu^+\nu_{\mu} \rightarrow
e^+\nu_e\bar{\nu}_{\mu}$~\cite{Stecker}, whereas tau neutrinos are produced
in the source only in negligible amounts. There are numerous different
models to estimate the neutrino spectrum from the observed photon
spectrum~\cite{SzaPro}, the average
neutrino energy being 1/20 of the primary proton energy.

In this paper we will consider particle acceleration in so-called parallel
shocks, i.e. the situation where the main component of the magnetic field
is parallel to the shock front velocity. At the shock front the magnetic
field is turbulent, which is crucial for the acceleration of the protons.
Outside the shock region the perturbations of the field are assumed to be
smaller. Although neutrinos can be produced in the turbulent region, in
most of the cases a neutrino observed at Earth propagates out of the
acceleration region not through the shock plane but through the region of
less turbulent magnetic field. We will make the assumption that the main
component of the magnetic field is homogenous and constant and that the
perturbations can be neglected. The effects of the magnetic field
turbulencies on neutrino propagation have been considered
in~\cite{DomotSahu}.  

In the case of Majorana neutrinos the magnetic interaction is described by
the Lagrangian
\bea{Lag}
{\cal L} = 
\frac{\mu_{ij}}{2} \bar\nu_i \sigma_{\alpha \beta} \nu_j^c F^{\alpha \beta}
\eea
where $\mu_{ij}$ is the magnetic transition moment of the interaction and
$F^{\alpha \beta}$ is the electromagnetic field strength tensor. This term
allows the flavour flip between $\nu_i$ and $\nu_j^c$. 
The equation of motion with magnetic transition in case of two neutrino
families, e.g. $\nu_\mu$ and ${\nu}_\tau$, is of the form~\cite{KimPevsner}
\bea{equmo}
i\left(\begin{array}{c} \dot{\nu}_\mu \\ \dot{\nu}^c_\tau \end{array}\right) 
= \frac{1}{2E}
\left(\begin{array}{cc} m_1^2 & 2E\mu B \\ 2E\mu B & m_2^2 \end{array}\right) 
\left(\begin{array}{c} {\nu}_\mu \\ \nu^c_\tau \end{array}\right),
\eea
where $B$ is the field transverse to the neutrino propagation.
For a magnetic transition to take place, the diagonal elements have to
be smaller than the off-diagonal elements: 
\bea{dequ}
\mu B \gg \frac{\Del_0}{2E},
\eea 
where $\Del_0=m_2^2 - m_1^2$. In case of blazars, by estimating
$B\simeq 1\
\rm G\simeq 2\times 10^{-2} \ \rm eV^2$, $\mu_B\approx 3\times 10^{-7}
\rm \ 1/eV$ and $E=10^{15}\ \rm eV$, this yields the condition
\bea{dem}
\Del_0 \ll 2EB \mu_B \frac{\mu}{\mu_B} 
\simeq 1.2 \times 10^7 \ {\rm eV^2} \ \frac{\mu}{\mu_B}\ \frac{E}{1\ \rm
PeV}\ \frac{B}{1\ \rm G}.
\eea 
Hence to be able to probe magnetic moment values of e.g. $10^{-12}
\mu_B$ the mass difference has to be $\Del_0 < 1.2\times 10^{-5}\ \rm
eV^2$. Notice that the smaller the mass difference, the stronger the
field or the higher the energy of the neutrino the smaller values of
magnetic moment can be tested.
If the mass difference $\Del_0$ obeys the condition of~\eq{dem}, the
Majorana neutrino flavor conversion probability in magnetic field is given by
\bea{conv}
P(\nu_{\rm L}\rightarrow\nu_{\rm R};r)=\sin^2 \left(\int_0^r \mu
B(r')\,{\rm d}r'\right).
\eea
The conversion is dependent only on neutrino magnetic moment $\mu$, magnetic 
field strength perpendicular to the neutrino propagation $B$ and 
the path length $r$ in the field. All neutrinos traversing the same distance
will thus undergo the same transition, independent of their energy.
\\

\noindent {\it 3. A simple model for neutrino flavor conversion in a
constant homogenous magnetic field.} In order to evaluate the conversion
probability~\eq{conv} one should have a model for the structure of the
magnetic field in the jet. As mentioned above, we assume a parallel shock
model for the acceleration where
the shock wave propagates along the magnetic field. This field is assumed
to be constant and homogenous in a circular region, perpendicular to the
field, and to drop quickly to zero outside that region. Inside the region
charged protons have a gyroradius $R=E_p/B$, which enlarges as the proton
gains energy in each crossing of the shock. All protons gyrate to the same
direction, and when a proton hits a gamma, a pion is produced. The
daughter neutrinos are born right after the collision, since the
pion and its longer living decay product muon stays in the region of
acceleration: with $E_{\mu}=10^{16} \ {\rm eV}$ the muon will fly
$\gamma \tau_{\mu} c \simeq 6\times 10^{10} \ {\rm m}$, which is much
less than the size of the accelerator (typically of the order of
$10^{14}\ {\rm m}$).  We will also assume that the photon number
density in different parts of the field is the same so that the pion
photoproduction (and neutrino production) rate is constant. 

Let $L$ denote the distance neutrino propagates in the direction
perpendicular to the magnetic field and towards the observer.
According to~\eq{conv} we can then write  
\bea{gamma}
P 
\simeq  
    \sin^2 \left( 8.7 \times 10^{12} \ \frac{\mu_{\nu}}{\mu_{\rm B}}\ 
\frac{B}{1\
\rm G}\ \frac{L}{10^{-2} \ \rm pc} \right) .
\eea
The typical values $B\simeq 1\,{\rm G}$ and $L\simeq 10^{-2} \ {\rm pc}$
correspond to a blazar jet near the central engine.

The effective distance $L$ depends on the site of creation of the
neutrino. As this is not known one has to calculate the average
conversion probability taking into account all possible path lengths.
This obviously depends on the energy of the neutrino, since the most
energetic neutrinos can be produced only at outskirts of the disk of
the jet cross section (or hot spot area), i.e.~where the gyroradius
$R=E_p/B$ of the primary proton is close to the radius $R_0$ of the
disk, while neutrinos with a lower energy can originate also deeper in
the disk. Following the geometry presented in Fig.~1, we can write the
average conversion probability for a given proton gyroradius $R$ in the
form:
\bea{Pave}
P_{\nu_e ,\nu_{\mu} \rightarrow\nu_{\tau} }^{\rm ave} = \frac{1}{\pi (R_0 -
R)^2} \int_0^{R_0-R} {\rm d}r 
\int_0^{2\pi} {\rm d} \phi\ r \ P_{\nu_e ,\nu_{\mu} \rightarrow\nu_{\tau} }
\left(\mu B L(r,\phi)\right).
\eea
Here $(r,\phi )$ is the location of the center point of the circular path
of the primary proton in polar coordinates with respect to the jet axis.
By simple geometry and algebra one can express the neutrino path length $L$ in 
terms of the polar coordiantes (see Fig.~1).

In Fig.~2 we have plotted the neutrino flavour ratio $\nu_\tau /\nu_{e,
\mu}=P_{\nu_{e,\mu} \rightarrow\nu_\tau}/(1-P_{\nu_{e,\mu}
\rightarrow\nu_\tau})$ as a function of energy for different magnetic
moment values. One can see that the tau neutrino component in the flux is
practically independent of the neutrino energy. At very high energies, 
however, when
gyroradius is close to the radius of the magnetic area, the ratio
clearly increases provided the magnetic moment is large enough. This is
because with increasing energy
the average distance
$L$ that neutrino propagates in the magnetic field decreases and becomes
close to the value favourable to a helicity transition. This can be seen
as an increase of the neutrino flavour ratio for the highest energies
achievable by the source. More important, however, is that with
certain  magnetic moment values the transition probability is higher
than 0.5 and consequently the flavour ratio is larger than 1. This is
the case e.g. for a blazar with
$R_0=10^{-2} \ {\rm pc}$ and magnetic field of $1 \,{\rm G}$ for
magnetic transition moment value $\mu \approx 1.4\times 10^{-13}\mu_B$.

We have plotted in Fig.~3 some examples of the neutrino flavour ratio
$\nu_\tau /\nu_{e, \mu}$ as a function
of magnetic transition moment at 1 PeV neutrino energy. We have used as
examples a blazar with $R_0=10^{-2} \ {\rm pc}$ and a magnetic field of
$1
\,{\rm G}$ and a hot spot with $R_0=1 \ {\rm kpc}$ and $0.1 \,{\rm mG}$.
For comparision, two other blazars are also represented, one
with $R_0=10 \ {\rm mpc}$ and with a weaker magnetic field, $0.1 \ {\rm
G}$, the other with a larger jet $R_0=1 \ {\rm pc}$ and $1 \ {\rm
G}$ field. The plot shows that the averaged ratio of the two flavours is
zero when the magnetic moment is too small to cause the conversion. It
increases from zero to  about 1.3 as the magnetic moment grows
one decade and finally settles slowly to 1. The region of the magnetic
moment value where the increase occurs depends on the characteristics
of the source, i.e. its size and magnetic field. 
One can see that the magnetic moment of the order of 
$10^{-13}\mu_B$ will
 already provide a flux with equal amounts of tau and muon
(or electron) neutrinos in a blazar with $R_0=10^{-2} \ {\rm pc}$ and
$B=1\ {\rm G}$.
In the case of hot spots one can test the magnetic moment down to values
around $10^{-14} \ \mu_B$. However, one has to remember that the real
distinction between the two cases is the mass difference of~\eq{dem}: with
the first example blazar it is $\Del_0\lsim 1.2\times 10^{-6}\ \rm eV^2$
and with the hot spot $\Del_0\lsim 1.2\times 10^{-11}\ \rm eV^2$. It 
is interesting to note that
the mass differences of these sizes have been used to explain the solar
neutrino problem via the MSW effect and vacuum oscillations~\cite{MSWvacua}, respectively. 

The recent results of Super-Kamiokande on atmospheric neutrinos~\cite{kamioka} can be explained by the existence of 
$\nu_\mu\leftrightarrow\nu_\tau$ or $\nu_\mu\leftrightarrow\nu_s$ oscillations ($\nu_s$ is a sterile
neutrino) with the oscillations parameters $\Delta m_\nu^2\simeq 10^{-3} - 10^{-2} \ {\rm eV}^2$ and $\sin^2 2\theta >0.8$. In the former case
there would be a $\nu_\tau$ component in the AGN neutrino flux which would shadow the effect we are considering. 
It should be emphasized, however, that if tau neutrinos are created by the $\nu_\mu\leftrightarrow\nu_\tau$ oscillations with the oscillation
parameters indicated by Super-Kamiokande results the flavour ratio would always average to 0.8~-~1. Thus the values of the 
$\nu_\tau /\nu_\mu$ larger than 1, or smaller than 0.8, could not be explained solely in terms of oscillations. 
In the case of $\nu_\mu\leftrightarrow\nu_s$ oscillations the appearance of $\nu_\tau$ would indicate magnetic transitions at AGN source, the oscillations affecting only the relative fluxes. Let us note that the constraints from the primordial nucleosynthesis seem to disfavour these oscillations in the Super-Kamiokande parameter range as these would increase the effective number of neutrino species~\cite{omat}. However, the present status of the limit on the number of neutrinos is somewhat unclear because of uncertainties in the $\rm D/ \, ^3H$ and $\rm ^4He$ abundance determinations~\cite{Gary}.
\\

\noindent{\it 4. Summary and discussion.} 
It is interesting to speculate about possible future data sets. Let
us suppose that we see $\nu_\tau$ from all extragalactic sources. If, as
believed, these neutrinos are not produced in the processes of the source,
this will then give a certain allowed area in the parameter space for both
neutrino oscillations and magnetic moments. If, however, we see $\tau$
neutrinos from some sources and not from all the other sources, we can rule out the
oscillations since the distance of all sources is of the same order of
magnitude (in astrophysical scales, more than 100 Mpc). In that case we
have been lucky to find the magnetic trasition moments of Majorana
neutrinos and confirmed the Majorana nature of neutrinos. In the mechanism 
we propose the ratio $\nu_\tau /\nu_{e,\mu}$ varies from zero to values somewhat
higher than one depending on the source as the transition
probability is different for different source parameters, i.e. magnetic field 
strength and size. This variation can be used as a signal to distinguish between
possible oscillations and magnetic transition moment as a reason to the appearance
of tau neutrino component in the flux.
One special case is to try to see difference in hot spot neutrino flux compared
to blazar neutrinos. Hot spot neutrinos have propagated in spatially large
but weak source fields, while the blazar neutrinos have propagated in 
much stronger but less extended fields. To see tau neutrinos
from hot spots but not from blazars would allow us to probe
magnetic moment down to values of the order of
$(10^{-15}-10^{-14})\mu_B$. Another example would be $\nu_\tau$ 
flux seen emanating from
blazars only, but not from hot spots. That would mean that the neutrino
mass difference of~\eq{dem} is greater than required for hot spot
neutrinos, of the order of $\Del_0\gsim 1.2\times 10^{-11}\ \rm eV^2$.
In these cases one has to distinguish between the
possible neutrino flux of the main engine and the hot spot. Finally, if
we see no $\nu_\tau$'s from any of these astrophysical sources one
can set an upper limit for the transition magnetic moments of the
order of
$(10^{-15}-10^{-14})\mu_B$. 
\\

\noindent{\it Acknowledgements.} One of the authors, P.K. would like to
thank Antti Sorri for useful discussions and the Vilho, Yrj\"o and 
Kalle V\"ais\"al\"a and Magnus Ehrnrooth foundations for financial support.
The work has been supported by the Academy of Finland under the contracts 
40677 and 10135224.


\pagebreak

\begin{figure}
\begin{center} \epsfxsize=7.5cm\epsfbox{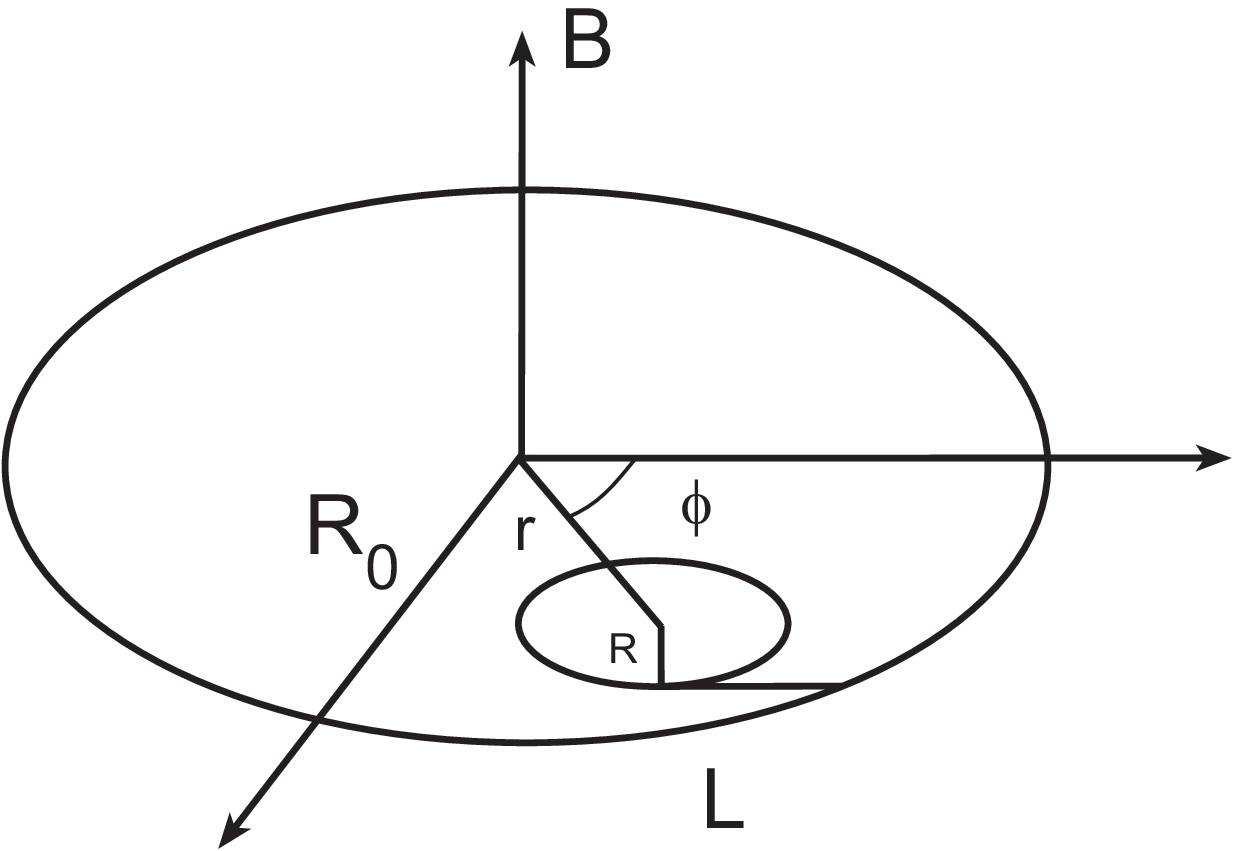} \end{center}
\caption{Geometry of the AGN. $R_0$ and $R$ are respectively the radius of the magnetized area and the gyroradius of the proton, ($r$,~$\phi$) is the center point of this gyration motion in polar coordinates and $L$ the neutrino path length in the magnetic field.}
\end{figure}

\begin{figure}
\begin{center} \epsfxsize=15.5cm\epsfbox{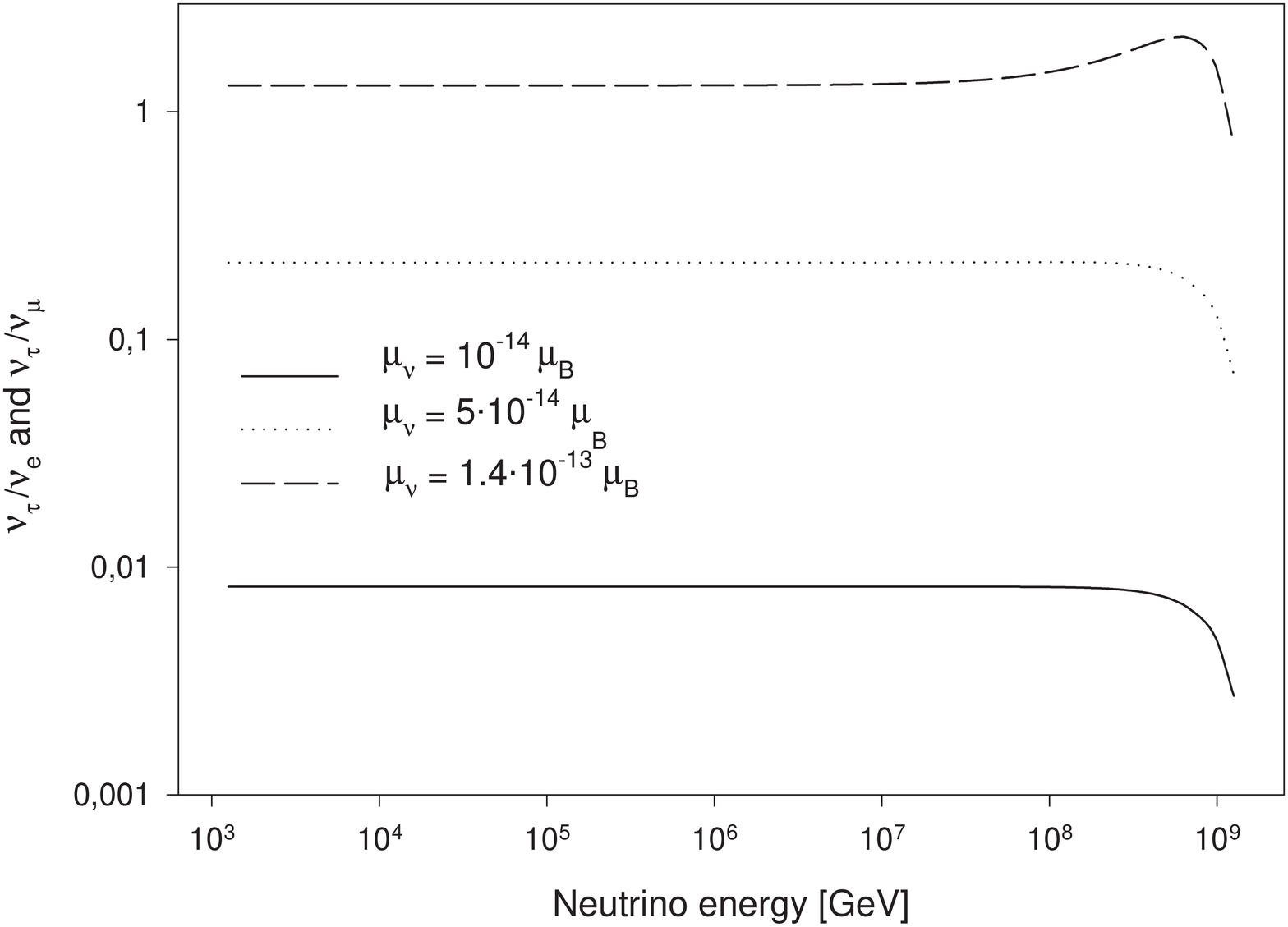} \end{center}
\caption{Neutrino flavour ratio $\nu_\tau /\nu_{e,\mu}$ as a function of energy for selected values of magnetic moment $\mu_\nu$. Here $R_0=10^{-2} \ {\rm pc}$ and magnetic field $1 \,{\rm G}$.}
\end{figure}

\begin{figure}
\begin{center} \epsfxsize=15.5cm\epsfbox{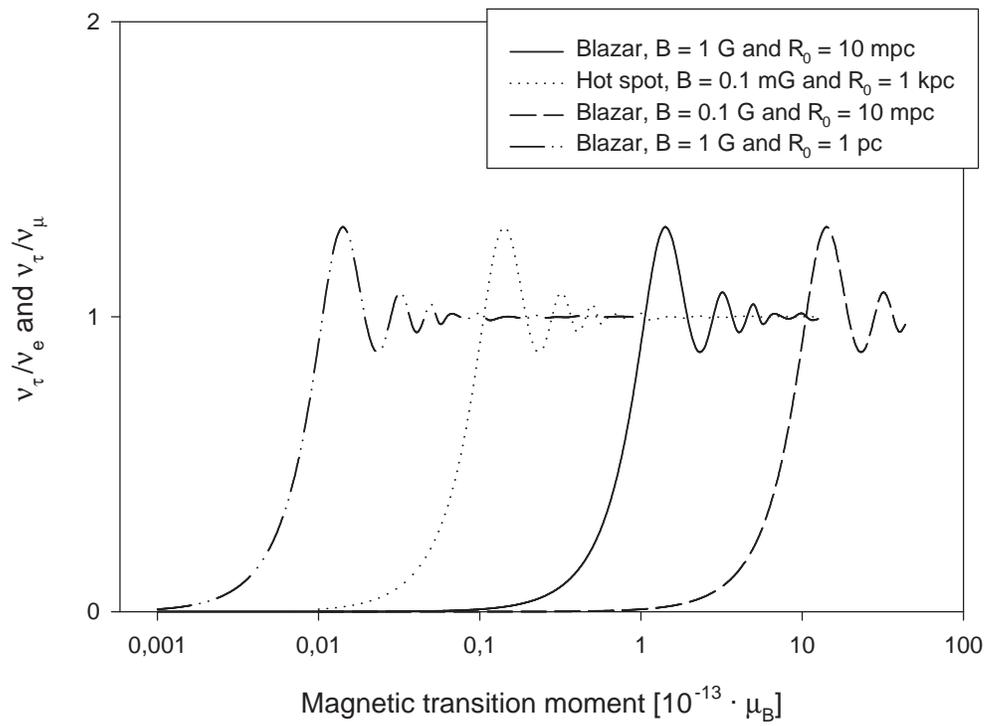} \end{center}
\caption{Neutrino flavour ratio $\nu_\tau /\nu_{e,\mu}$ from different sources.}
\end{figure}

\end{document}